\documentclass[11pt,a4paper]{article}

\usepackage[margin=1in]{geometry}
\usepackage[T1]{fontenc}
\usepackage[utf8]{inputenc}
\usepackage{amsmath}
\usepackage{amssymb}
\usepackage{graphicx}
\usepackage{booktabs}
\usepackage{siunitx}
\usepackage{multirow}
\usepackage{subcaption}
\usepackage{ragged2e}
\usepackage{xcolor}
\usepackage{hyperref}

\hypersetup{
  colorlinks=true,
  linkcolor=blue!50!black,
  citecolor=blue!50!black,
  urlcolor=blue!50!black
}

\title{Patagium and tail morphology shape aerodynamic performance and control authority in gliding-mammal-inspired wings}

\author{
\parbox{0.92\textwidth}{\centering
Liming Zheng\textsuperscript{1,*},
Baihui Chen\textsuperscript{2}\\
Alexander van Zuijlen\textsuperscript{2}
and Salua Hamaza\textsuperscript{1}\\[0.65em]
\footnotesize \textsuperscript{1}BioMorphic Intelligence Lab, Department of Control and Operations,\\
Faculty of Aerospace Engineering, Delft University of Technology, Delft, The Netherlands\\[0.25em]
\footnotesize \textsuperscript{2}Aerodynamics, Department of Flow Physics and Technology,\\
Faculty of Aerospace Engineering, Delft University of Technology, Delft, The Netherlands\\[0.25em]
\footnotesize \textsuperscript{*}Corresponding author: \href{mailto:l.zheng-1@tudelft.nl}{l.zheng-1@tudelft.nl}
}
}

\date{}

\begin{document}
\justifying
\maketitle

\begin{center}
\small Preprint version. Manuscript under review at \textit{Bioinspiration \& Biomimetics}.
\end{center}

\begin{abstract}
Gliding mammals exhibit diverse patagium and tail/uropatagium morphologies that may influence aerodynamic performance and maneuverability. Here, we use computational fluid dynamics to isolate the aerodynamic effects of representative gliding-mammal-inspired morphologies under controlled flow conditions. Three patagium configurations were compared to evaluate the effects of membrane outline on lift generation, drag, stall behavior and pitching moment. Three tail/uropatagium configurations were further tested under baseline, symmetric-deflection and asymmetric-deflection conditions to assess their longitudinal and lateral control authority. The results show that a broader patagium configuration generated the highest lift and lift coefficient, whereas an intermediate patagium morphology showed a smoother post-stall response with lower drag. For the tail configurations, the colugo-like integrated uropatagium enhanced lift and pitch-control authority under symmetric deflection, while the flat-tail configuration produced stronger rolling and yawing responses under asymmetric deflection. These findings indicate that gliding-mammal-inspired morphologies produce distinct aerodynamic trade-offs rather than a single optimal design. The results provide insight into the functional diversity of gliding mammal morphology and offer design guidance for bioinspired morphing aerial robots.
\end{abstract}

\noindent\textbf{Keywords:} gliding mammals; bioinspired aerodynamics; patagium; uropatagium; tail morphology; computational fluid dynamics; morphing aerial robots

\section{Introduction}

Gliding mammals have evolved diverse membrane-based morphologies that allow them to perform controlled aerial locomotion without powered flight\cite{bookGlidingMammals}. Unlike birds\cite{robot-Lentink-sr1,robot-Lentink-sr2,robot-Dario-sr1,robot-Dario-sr2} and bats\cite{science_bat}, which generate thrust through active flapping, gliding mammals rely primarily on gravitational potential energy and aerodynamic forces generated by their patagia. During a glide, the limbs spread the membrane to form an effective lifting surface, while the body, tail and uropatagium contribute to attitude control and maneuvering\cite{bio-non-equilibrium-gliding}. This form of locomotion is found in several mammalian groups, including flying squirrels\cite{robot-sr2025-squirrel}, gliding marsupials, scaly-tailed flying squirrels and colugos, which exhibit substantial variation in body size, patagium outline and tail morphology.

The patagium is the primary aerodynamic surface of gliding mammals\cite{bookGlidingMammals,beihang-ieee-access}. Its outline determines the effective planform, projected area and spatial distribution of aerodynamic loading. Previous biological and aerodynamic studies have shown that gliding performance is affected by body posture, membrane area, wing loading and glide angle\cite{robot-nc-Liming}. However, the aerodynamic role of different patagium outlines remains difficult to isolate in animals because morphology, posture, body size and behavior vary simultaneously among species~\cite{bookGlidingMammals}. A controlled comparative analysis is therefore useful for understanding how changes in membrane geometry alone can influence aerodynamic performance.

\begin{figure}
 \centering
 \includegraphics[width=0.9\textwidth]{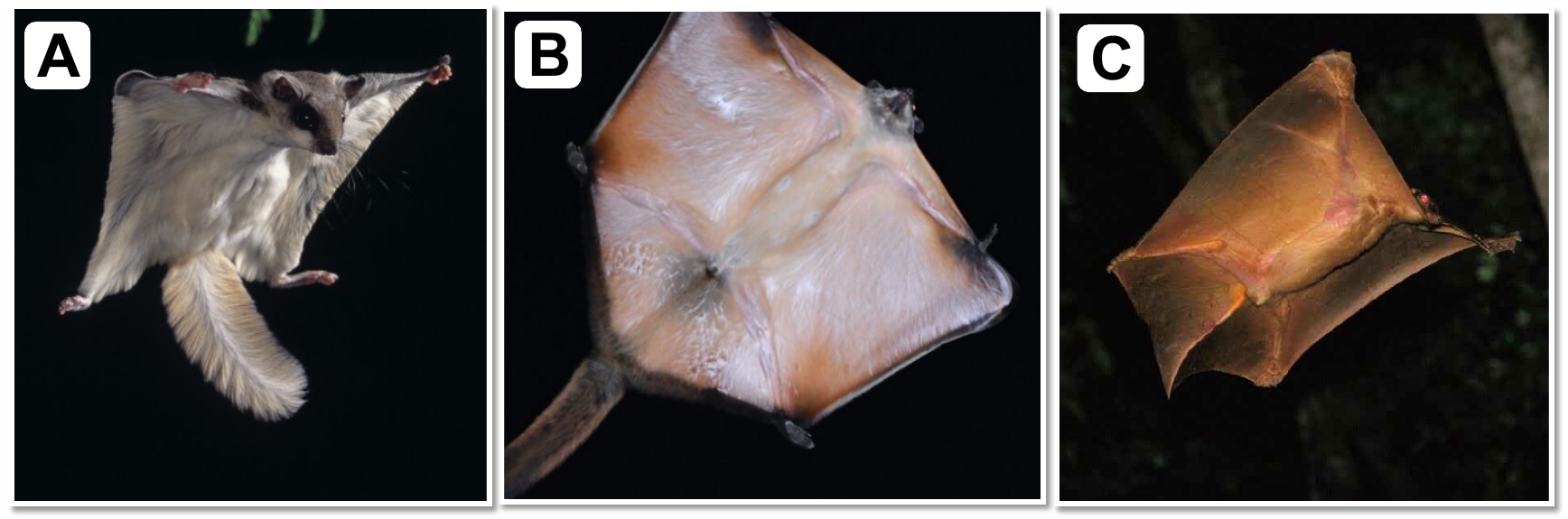}
 \caption{Representative gliding mammals illustrating the diversity of patagium morphology. 
(A) \textit{Northern flying squirrel}. 
(B) \textit{Red giant flying squirrel}. 
(C) \textit{Colugo}. 
These representative species motivated the selection of three gliding-mammal-inspired patagium configurations used in the comparative aerodynamic analysis.}
\label{fig1}
\end{figure}

In addition to the patagium, tail and uropatagium morphology may play an important role in gliding control. In some species, such as many flying squirrels, the tail is a distinct structure that can potentially contribute to lateral maneuvering and attitude adjustment~\cite{bio-red-giant-fs,bio-fs-high-aoa} (Figure~\ref{fig1}A and B). In colugos~\cite{bookGlidingMammals,colugo1,colugo2}, the tail is largely enclosed within the uropatagium, forming a more continuous posterior membrane (Figure~\ref{fig1}C). These different morphologies suggest that the tail and uropatagium may provide different aerodynamic functions. A separated tail may act as a control surface for roll or yaw, whereas an integrated uropatagium may increase lifting area and pitch-control authority. However, the aerodynamic trade-offs among these configurations remain insufficiently quantified.

Understanding these morphological effects is also relevant for bioinspired aerial robots~\cite{book-CLCD}. Small gliding robots and morphing aerial platforms~\cite{robot-Lentink-sr2,robot-Dario-sr2,morphing_wing_xr} often operate under strong constraints on mass, actuation and control authority. Inspired by gliding mammals, membrane wings and morphing tails can offer lightweight solutions for enhanced stability, agility and maneuverability~\cite{robot-nc-Liming}. However, the design of such systems requires an understanding of how morphology affects aerodynamic force and moment generation. In particular, it is important to determine whether different patagium and tail morphologies favor lift, stall robustness, pitch control or lateral maneuverability~\cite{bio-fs-perching-trajectory}.

Here, we use computational fluid dynamics (CFD) to compare representative gliding-mammal-inspired patagium and tail/uropatagium morphologies under controlled conditions. The study consists of two parts. First, we compare three simplified patagium configurations with different membrane outlines while keeping the main structural dimensions fixed. The aerodynamic performance of these configurations is evaluated using lift force, lift coefficient, drag coefficient, pitching moment coefficient and pressure distributions. Second, we compare three representative tail/uropatagium configurations motivated by morphological variation among gliding mammals. These configurations are tested under baseline, symmetric-deflection and asymmetric-deflection conditions to evaluate their effects on lift, pitch, roll and yaw coefficients.

The purpose of this study is not to reconstruct the full anatomy or flight behavior of any individual species. Instead, the aim is to isolate first-order aerodynamic effects of selected morphological features and identify the trade-offs associated with different gliding-mammal-inspired designs. By comparing patagium outlines and tail/uropatagium configurations in a unified computational framework, this work provides insight into how morphology can shape aerodynamic performance and control authority in both gliding mammals and bioinspired morphing aerial robots.

\section{Morphological design and computational methods}

\subsection{Selection of representative morphologies}

The aerodynamic effects of gliding-mammal-inspired morphology were investigated through two sets of simplified three-dimensional models. The first set was designed to isolate the effect of patagium outline, while the second set was designed to isolate the effect of tail and uropatagium morphology. In both cases, the goal was not to reconstruct a specific animal in full anatomical detail, but to compare representative morphological features under controlled aerodynamic conditions.

For the patagium study, three wing configurations were selected and denoted as W-1, W-2 and W-3. These configurations were inspired by representative gliding mammals with different patagium outlines (Figure~\ref{fig1} and Figure~\ref{fig3}). To allow direct comparison, the three models were designed with the same limb length and sweep angle. The membrane was modeled as a flat rigid surface, while the tail and uropatagium were omitted. This rigid-membrane simplification was adopted to isolate the aerodynamic effect of planform geometry from membrane compliance and fluid--structure interaction. Therefore, the only major geometric variable in this comparison was the patagium outline. This simplification allowed the aerodynamic influence of membrane shape to be evaluated independently from the effects of tail morphology and membrane deformation.

\begin{figure}
 \centering
 \includegraphics[width=0.9\textwidth]{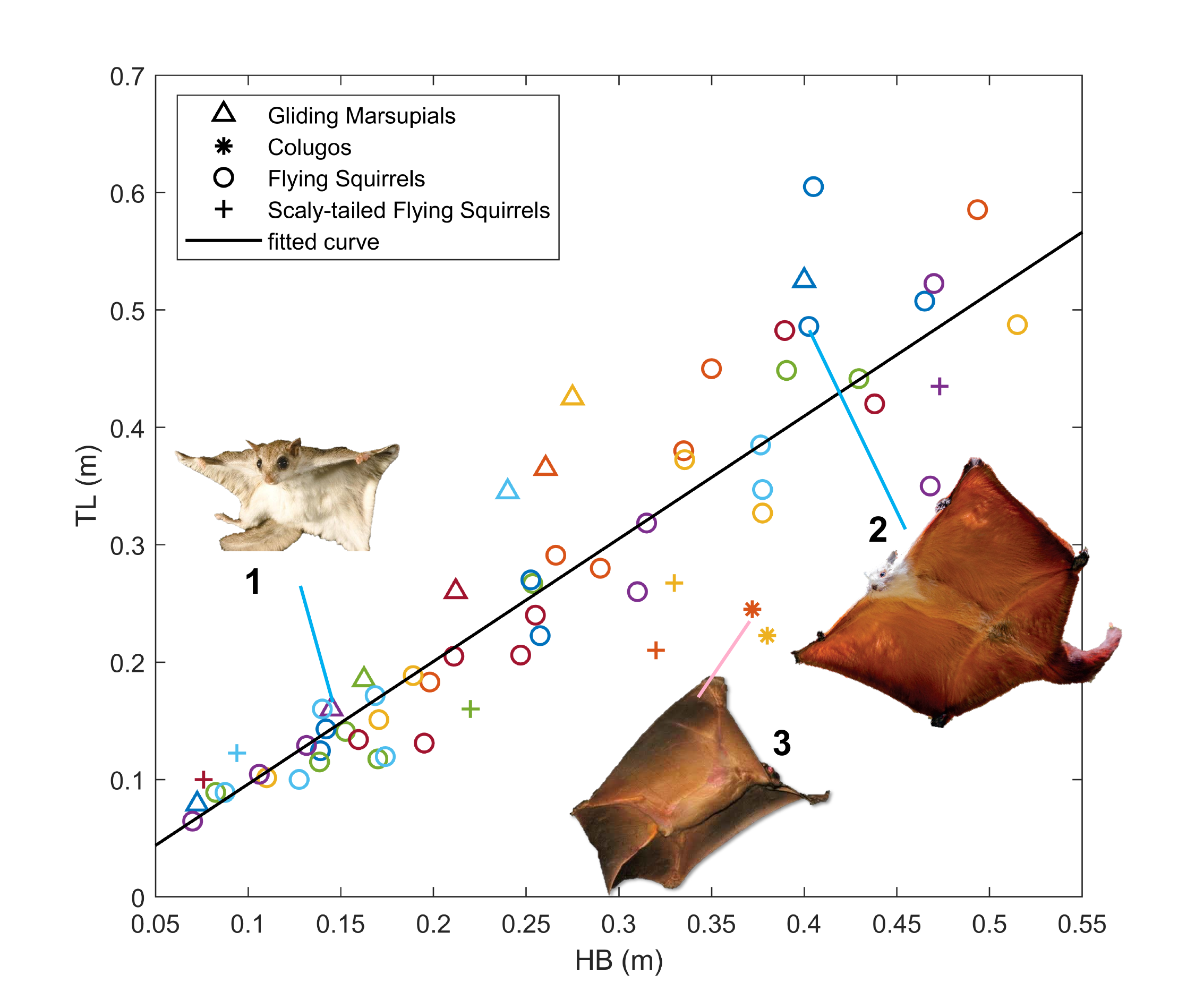}
 \caption{Morphological diversity of tail length among gliding mammals. 
Tail length (TL) is plotted against head--body length (HB) for 53 species of gliding mammals, including gliding marsupials, colugos, flying squirrels and scaly-tailed flying squirrels. 
The solid black line indicates a linear fit to the collected morphological data. 
The representative species highlighted in the plot are: 1, northern flying squirrel; 2, red giant flying squirrel; and 3, colugo. 
These data motivated the selection and scaling of the tail/uropatagium configurations used in the subsequent aerodynamic simulations.}
\label{fig5}
\end{figure}

For the tail and uropatagium study, the W-3 patagium configuration was used as the baseline wing because it generated the highest lift among the three patagium configurations. Three representative tail/uropatagium morphologies were then defined. T1 represents a flat-tail configuration without a fully developed uropatagium, inspired by flying squirrels. T2 represents an intermediate configuration with an elongated tail and partial uropatagium, inspired by the red giant flying squirrel. T3 represents a colugo-like configuration in which the tail is enclosed within the uropatagium. The tail lengths of T1 and T2 were determined based on a linear fit between tail length (TL) and head--body length (HB) collected from 53 species of gliding mammals, while T3 was scaled according to the representative colugo morphology shown in Figure~\ref{fig5}.

Each tail morphology was simulated under three deflection conditions (Figure~\ref{fig6}A). The baseline condition, denoted as ``a'', corresponds to zero tail deflection, with \(\delta_e = 0^\circ\) and \(\delta_a = 0^\circ\). The symmetric-deflection condition, denoted as ``b'', corresponds to an elevator-like deflection, with \(\delta_e = 15^\circ\) and \(\delta_a = 0^\circ\). The asymmetric-deflection condition, denoted as ``c'', corresponds to a lateral-control deflection, with \(\delta_e = 15^\circ\) and \(\delta_a = 25^\circ\). This resulted in nine tail configurations: T1-a, T1-b, T1-c, T2-a, T2-b, T2-c, T3-a, T3-b and T3-c. Here, \(\delta_e\) represents the symmetric tail deflection angle, while \(\delta_a\) represents the asymmetric deflection angle.

\subsection{CFD setup}

All aerodynamic simulations were performed using Ansys Fluent 2020 R2. The computational domain was a rectangular box with dimensions of \(3~\mathrm{m} \times 3~\mathrm{m} \times 4~\mathrm{m}\), with the gliding-mammal-inspired model placed at the centre of the domain. Equivalently, the domain size was chosen to be sufficiently large relative to the model length to reduce boundary effects. One symmetry plane was used, while the remaining outer boundaries were prescribed as velocity-inlet boundaries~\cite{cfd-1}. The model surface was treated as a no-slip wall.

\begin{figure}
 \centering
 \includegraphics[width=0.99\textwidth]{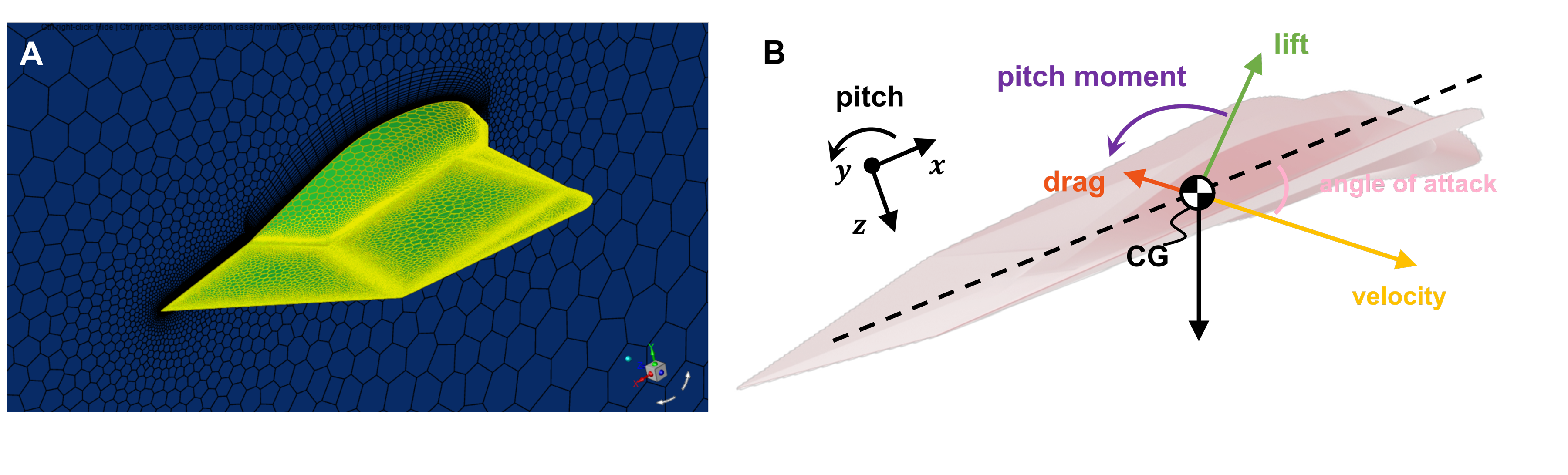}
 \caption{Computational setup and aerodynamic definitions used in the CFD simulations. 
(A) Representative near-body mesh around the gliding-mammal-inspired configuration, showing the surface mesh and boundary-layer refinement. 
(B) Definition of the body reference line, angle of attack, aerodynamic forces and pitching moment. 
The angle of attack, $\alpha$, was defined as the angle between the incoming velocity vector and the body reference line from the anterior body point to the posterior connection point between the body, membrane and tail. 
Lift and drag were defined relative to the incoming flow direction, while the pitching moment was evaluated about the prescribed moment reference point.}
\label{fig2}
\end{figure}

The free-stream velocity was set to \(U_\infty = 12~\mathrm{m~s^{-1}}\), corresponding to a Reynolds number of approximately \(2.4 \times 10^5\). The air density was set to \(\rho = 1.205~\mathrm{kg~m^{-3}}\). The angle of attack, \(\alpha\), was defined as the angle between the incoming velocity vector and the body reference line connecting the anterior body point to the posterior connection point between the body, membrane and tail (Figure~\ref{fig2}B). For each configuration, simulations were conducted over an angle-of-attack range from \(-2^\circ\) to \(60^\circ\), with a step size of \(2^\circ\).

The geometries were discretized using an hexahedral-dominant unstructured mesh with boundary-layer refinement near the model surface (Figure~\ref{fig2}A). The mesh contained approximately \(3.0 \times 10^6\) cells. The boundary-layer mesh was generated with a first-layer height of \(6.5 \times 10^{-5}~\mathrm{m}\), corresponding to a target \(y^+ \approx 1\), with 30 layers and a growth rate of 1.15. The same meshing strategy was used for all configurations to ensure consistency across the comparative simulations.

The simulations used the realizable \(k\)-\(\varepsilon\) turbulence model~\cite{cfd-3,cfd-4}. The residual convergence criterion for continuity was set below \(10^{-5}\). Surface static pressure distributions and local velocity vectors were post-processed at \(\alpha = 20^\circ\) to visualize the flow-field differences among the patagium configurations.

\subsection{Aerodynamic coefficients}

The aerodynamic forces and moments were extracted from the CFD simulations for each angle of attack. Lift \(L\) and drag \(D\) were defined relative to the incoming free-stream direction (Figure~\ref{fig2}B). The lift coefficient and drag coefficient were calculated as
\begin{equation}
C_L = \frac{L}{\frac{1}{2}\rho U_\infty^2 S},
\end{equation}
and
\begin{equation}
C_D = \frac{D}{\frac{1}{2}\rho U_\infty^2 S},
\end{equation}
where \(\rho\) is the air density, \(U_\infty\) is the free-stream velocity and \(S\) is the reference area.

The pitching, rolling and yawing moment coefficients were calculated as
\begin{equation}
C_m = \frac{M_y}{\frac{1}{2}\rho U_\infty^2 S c},
\end{equation}
\begin{equation}
C_l = \frac{M_x}{\frac{1}{2}\rho U_\infty^2 S b},
\end{equation}
and
\begin{equation}
C_n = \frac{M_z}{\frac{1}{2}\rho U_\infty^2 S b},
\end{equation}
where \(M_y\), \(M_x\) and \(M_z\) are the pitching, rolling and yawing moments, respectively. The reference chord and reference span are denoted by \(c\) and \(b\). The moment coefficients were evaluated about the prescribed moment reference point shown in Figure~\ref{fig2}B. The same reference point was used for all configurations to enable relative comparison of moment trends among morphologies.

For the patagium comparison, lift force, \(C_L\), \(C_D\) and \(C_m\) were used to evaluate lift generation, drag penalty and longitudinal pitching characteristics. For the tail and uropatagium comparison, \(C_L\) and \(C_m\) were used to evaluate the longitudinal effects of symmetric tail deflection, while \(C_l\) and \(C_n\) were used to evaluate the lateral-control effects of asymmetric tail deflection.

\section{Results}

To evaluate the aerodynamic effects of gliding-mammal-inspired morphologies, we conducted two sets of comparative CFD simulations. The first set focused on patagium morphology, where three wing configurations with different membrane outlines were compared under the same flow conditions. The second set focused on tail and uropatagium morphology, where three representative tail configurations were tested under baseline, symmetric-deflection and asymmetric-deflection conditions. The results are presented in terms of aerodynamic force and moment coefficients, together with pressure and flow-field visualizations.

\subsection{Aerodynamic effects of patagium morphology}

\begin{figure}
 \centering
 \includegraphics[width=0.99\textwidth]{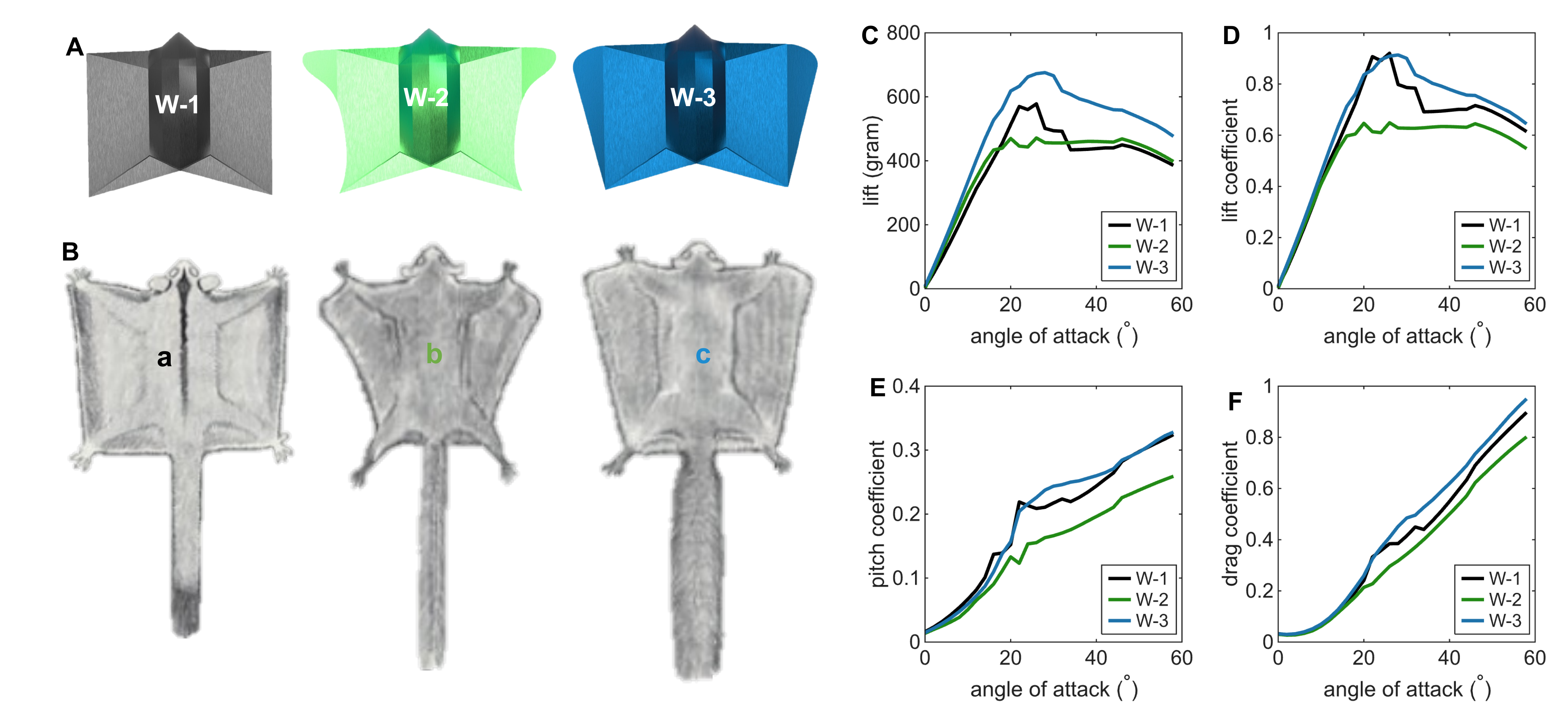}
 \caption{Influence of patagium morphology on aerodynamic performance. 
(A) Three gliding-mammal-inspired patagium configurations used in the CFD simulations: W-1, W-2 and W-3. 
(B) Representative biological morphologies motivating the three configurations. 
(C) Lift force as a function of angle of attack. 
(D) Lift coefficient, $C_L$, as a function of angle of attack. 
(E) Pitching moment coefficient, $C_m$, as a function of angle of attack. 
(F) Drag coefficient, $C_D$, as a function of angle of attack. 
W-3 generated the highest lift and lift coefficient over most of the tested angle-of-attack range, whereas W-2 showed a smoother post-stall response with lower overall lift generation.}
\label{fig3}
\end{figure}

We first compared three simplified patagium configurations, denoted as W-1, W-2 and W-3 (Figure~\ref{fig3}A). These configurations were designed to represent different gliding-mammal-inspired membrane outlines while keeping the main geometric parameters, such as limb length and sweep angle, consistent. Representative biological morphologies are shown in Figure~\ref{fig3}B. This comparison was intended to isolate the first-order aerodynamic effect of patagium outline under controlled computational conditions.

The lift force increased with angle of attack for all three configurations at low and moderate angles, followed by a peak and a subsequent decrease at higher angles of attack (Figure~\ref{fig3}C). Among the three configurations, W-3 generated the largest lift over most of the tested angle-of-attack range. Its lift increased rapidly at low angles of attack and reached the highest peak value among the three designs. W-1 produced intermediate lift, whereas W-2 generated the lowest lift over most of the range.

Although W-2 produced lower lift, its lift curve showed a smoother variation after stall. In contrast, W-1 and W-3 showed more pronounced lift peaks followed by clearer reductions at higher angles of attack. This indicates that the patagium outline influences not only maximum lift generation, but also the way aerodynamic performance degrades after stall. Therefore, the three configurations should not be interpreted simply as better or worse designs, but as different aerodynamic trade-offs.

The lift coefficient, \(C_L\), showed similar trends to the lift force (Figure~\ref{fig3}D). W-3 reached the highest \(C_L\), indicating that its stronger lift generation was not only a consequence of projected area, but was also associated with the aerodynamic loading produced by its membrane shape. W-1 reached a comparable peak at moderate angles of attack, but its coefficient decreased more noticeably at higher angles. W-2 showed a lower \(C_L\) throughout most of the tested range, while maintaining a comparatively smooth post-stall response.

The drag coefficient, \(C_D\), increased with angle of attack for all three configurations (Figure~\ref{fig3}F). W-3 generally produced higher drag at moderate and high angles of attack, consistent with its stronger lift generation and larger aerodynamic loading. W-2 showed a lower drag coefficient over much of the tested range, indicating that its reduced lift generation was accompanied by a lower drag penalty. W-1 again showed intermediate behavior. These results suggest a lift--drag trade-off among the three patagium morphologies.

The pitching moment coefficient, \(C_m\), also varied with patagium morphology (Figure~\ref{fig3}E). This indicates that changes in membrane outline modify not only the total aerodynamic force, but also the distribution of this force relative to the body and reference point. Configurations with stronger lift generation also showed larger changes in pitching moment at moderate and high angles of attack. Thus, patagium morphology can influence both aerodynamic performance and longitudinal pitching characteristics.

\begin{figure}
 \centering
 \includegraphics[width=0.99\textwidth]{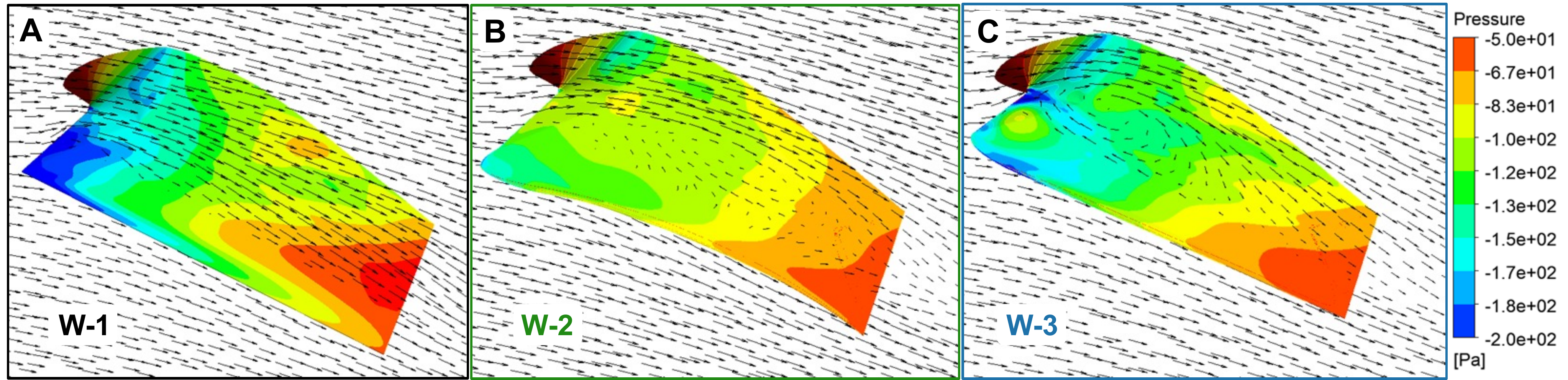}
 \caption{Surface static pressure and flow-field distributions around the three patagium configurations. 
Surface static pressure distributions and local velocity vectors are shown for (A) W-1, (B) W-2 and (C) W-3 at an angle of attack of $\alpha = 20^\circ$. 
The pressure contour represents static pressure, and the arrows indicate the local flow direction around the membrane surface. 
This visualization provides a qualitative explanation for the aerodynamic trends shown in Figure~\ref{fig3}, highlighting how changes in patagium geometry modify the pressure distribution and local flow organization. 
W-3 exhibits a stronger low-pressure region over the membrane surface, consistent with its higher lift coefficient, whereas W-2 shows a more distributed pressure pattern associated with a smoother post-stall response.}
 \label{fig4}
\end{figure}

To further interpret these coefficient trends, we examined the surface pressure distribution and local flow direction around the three configurations at a representative angle of attack (Figure~\ref{fig4}). W-3 exhibited a stronger low-pressure region near the leading-edge portion of the membrane (Figure~\ref{fig4}C), consistent with its higher lift and lift coefficient in Figure~\ref{fig3}. The broader membrane distribution of W-3 allows a larger portion of the wing surface to contribute to lift generation.

In comparison, W-2 showed a more spatially distributed pressure pattern (Figure~\ref{fig4}B). The magnitude of the low-pressure region was less pronounced than in W-3, consistent with the lower lift coefficient shown in Figure~\ref{fig3}D. However, this more uniform pressure distribution may also be related to the smoother post-stall behavior observed in the lift curve. W-1 showed an intermediate pressure distribution, with localized low-pressure regions and aerodynamic coefficients between W-2 and W-3.

The local flow vectors in Figure~\ref{fig4} also show that the three patagium geometries modify the flow direction over the membrane surface. Although the present simulations do not include flexible membrane deformation, these flow-field visualizations indicate that the membrane outline alone can substantially alter pressure loading and local flow organization. Together, Figure~\ref{fig3} and Figure~\ref{fig4} show that patagium morphology affects lift generation, drag, stall behavior and pitching characteristics through geometry-induced changes in aerodynamic loading.

\subsection{Aerodynamic effects of tail and uropatagium morphology}

We next evaluated the aerodynamic effects of tail and uropatagium morphology. To motivate the selection of representative configurations, we first compared tail length (TL) and head--body length (HB) across 53 species of gliding mammals, including gliding marsupials, colugos, flying squirrels and scaly-tailed flying squirrels (Figure~\ref{fig5}). The collected data show a positive relationship between TL and HB, indicating that larger gliding mammals generally tend to have longer tails. However, the data also show considerable variation around the fitted trend line, suggesting that tail morphology is not determined by body size alone.

The representative species highlighted in Figure~\ref{fig5} illustrate this variation. Some gliding mammals have a relatively distinct and flat tail, while others possess a more integrated uropatagium in which the tail is partly or largely enclosed by the membrane. Based on this morphological diversity, three simplified tail/uropatagium configurations were selected for aerodynamic comparison. T1 represents a flat-tail configuration without a fully developed uropatagium. T2 represents an intermediate morphology with an elongated tail and partial uropatagium. T3 represents a colugo-like morphology in which the tail is enclosed within the uropatagium.

\begin{figure}
 \centering
 \includegraphics[width=0.99\textwidth]{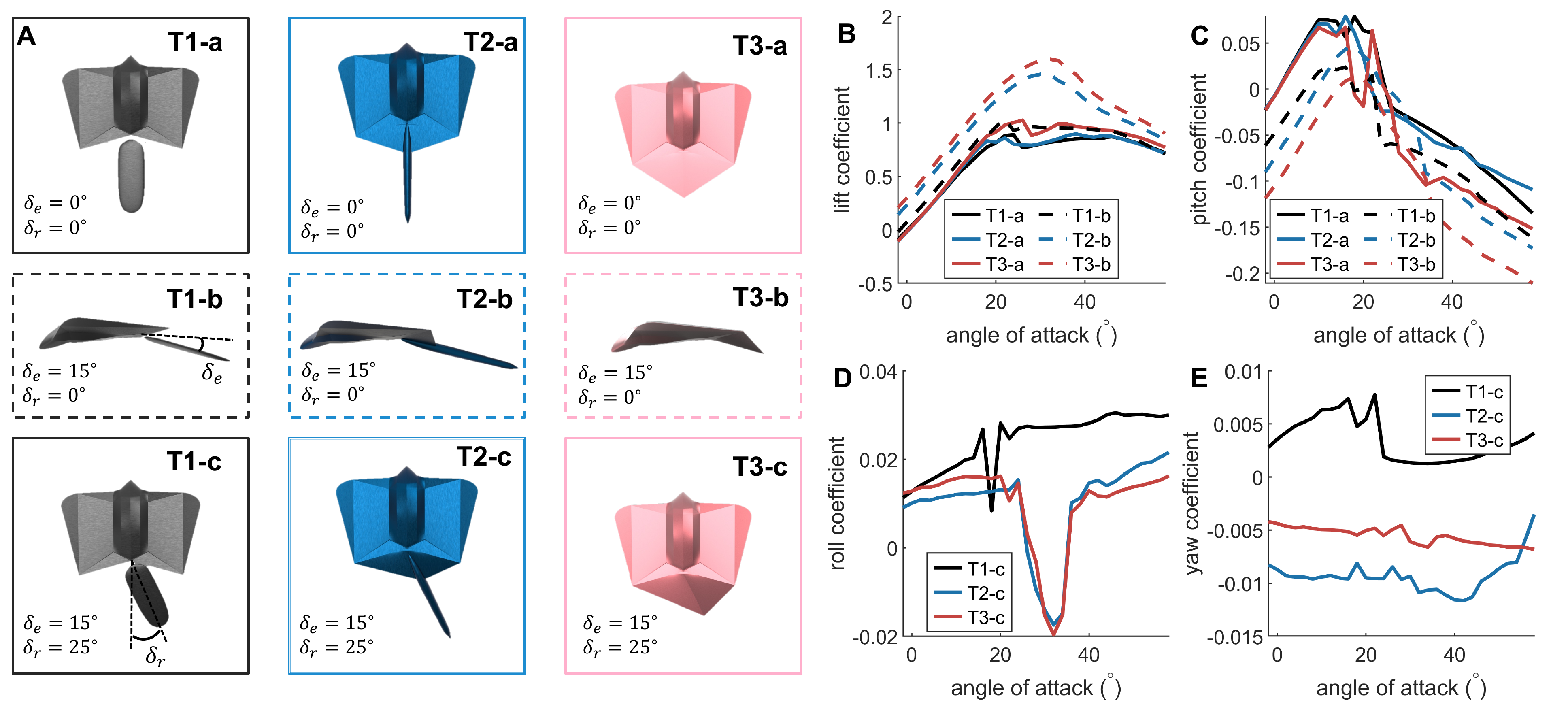}
 \caption{Influence of tail and uropatagium morphology on aerodynamic force and moment coefficients. 
(A) Three tail/uropatagium configurations inspired by representative gliding mammals were simulated under three deflection conditions. 
T1 represents a flat-tail configuration without a uropatagium, T2 represents a partial-uropatagium configuration with an elongated tail, and T3 represents a colugo-like configuration in which the tail is enclosed within the uropatagium. 
The baseline cases T1-a, T2-a and T3-a have $\delta_e = 0^\circ$ and $\delta_a = 0^\circ$; the symmetric-deflection cases T1-b, T2-b and T3-b have $\delta_e = 15^\circ$ and $\delta_a = 0^\circ$; and the asymmetric-deflection cases T1-c, T2-c and T3-c have $\delta_e = 15^\circ$ and $\delta_a = 25^\circ$. 
(B) Lift coefficient, $C_L$, and (C) pitching moment coefficient, $C_m$, for the baseline and symmetric-deflection cases. 
(D) Rolling moment coefficient, $C_l$, and (E) yawing moment coefficient, $C_n$, for the asymmetric-deflection cases. 
The results show that the colugo-like configuration provides the strongest lift enhancement and pitch control authority under symmetric deflection, whereas the flat-tail configuration produces larger lateral-control responses under asymmetric deflection.}
\label{fig6}
\end{figure}

The three configurations were simulated under three deflection conditions (Figure~\ref{fig6}A). The baseline cases, T1-a, T2-a and T3-a, correspond to zero tail deflection. The symmetric-deflection cases, T1-b, T2-b and T3-b, represent an elevator-like deflection. The asymmetric-deflection cases, T1-c, T2-c and T3-c, represent a lateral-control condition. This setup allowed us to compare the effects of tail morphology on both longitudinal and lateral aerodynamic control.

For the baseline and symmetric-deflection cases, all three configurations showed increasing lift coefficient with angle of attack before reaching a peak and entering the post-stall regime (Figure~\ref{fig6}B). The effect of symmetric deflection differed substantially among the three morphologies. The colugo-like configuration T3 showed the strongest lift enhancement under symmetric deflection. In particular, T3-b generated a higher \(C_L\) than its baseline counterpart over a broad range of angles of attack. This indicates that an integrated uropatagium can function as an effective posterior lifting surface when deflected.

T1 and T2 showed smaller lift enhancement under the same deflection condition. For T1, the flat tail is more separated from the main patagium and therefore contributes less to the effective lifting surface. T2 showed an intermediate response, consistent with its intermediate geometry between the flat-tail and colugo-like configurations. These results suggest that the degree of integration between the tail and uropatagium affects how effectively tail deflection modifies lift.

The pitching moment coefficient was also strongly affected by tail morphology and deflection state (Figure~\ref{fig6}C). Symmetric deflection produced noticeable changes in pitching moment for all configurations, but the magnitude of this effect differed among T1, T2 and T3. T3 produced the strongest pitch response, suggesting higher longitudinal control authority. This is consistent with the larger posterior membrane surface of the colugo-like uropatagium, which can shift aerodynamic loading and generate a stronger pitching moment. T1 showed a weaker pitch response, while T2 again exhibited intermediate behavior.

For the asymmetric-deflection cases, the rolling and yawing moment coefficients revealed different lateral-control characteristics (Figure~\ref{fig6}D,E). The flat-tail configuration T1 produced the largest rolling moment response among the three configurations. This suggests that although T1 is less effective for lift enhancement and pitch control, it can generate stronger lateral aerodynamic moments under asymmetric deflection. Such a response may be beneficial for roll control or turning maneuvers.

The yawing moment coefficient also differed among the configurations (Figure~\ref{fig6}E). T1 and T2 showed more pronounced yaw responses than T3 over parts of the tested angle-of-attack range. The reduced lateral response of T3 may be related to the more symmetric integration of the tail within the uropatagium, which enhances lift and pitch authority but reduces the distinct side-force contribution of a separated tail. In contrast, the more exposed tail geometries of T1 and T2 can generate larger asymmetric aerodynamic effects when deflected.

These results indicate a trade-off among the tail and uropatagium morphologies. The colugo-like T3 configuration is advantageous for lift enhancement and pitch control, whereas the flat-tail T1 configuration provides stronger lateral-control authority. The intermediate T2 configuration shows aerodynamic behavior between these two extremes. Therefore, tail and uropatagium morphology can affect not only aerodynamic force generation, but also the type of control authority available to the gliding body.

\section{Discussion and conclusions}

The present study used CFD simulations to compare how simplified gliding-mammal-inspired patagium and tail/uropatagium morphologies affect aerodynamic performance and control authority. The results show that variations in membrane outline and tail integration can produce distinct aerodynamic consequences under identical flow conditions. Rather than identifying a single morphology that is universally superior, the simulations reveal trade-offs among lift generation, drag, stall behavior, pitching moment and lateral-control authority.

For the patagium configurations, W-3 generated the highest lift and lift coefficient over most of the tested angle-of-attack range. The pressure distribution suggests that this improvement is associated with a stronger low-pressure region over the membrane surface. However, W-3 also produced higher drag and stronger changes in pitching moment, particularly at moderate and high angles of attack. In contrast, W-2 produced lower lift and drag, but showed a smoother post-stall response. These results suggest that patagium shape can tune the balance between peak aerodynamic force and aerodynamic robustness after stall.

This trade-off may be relevant for understanding the diversity of gliding mammal morphologies. In natural gliding, animals do not optimize for lift alone. They must also maintain stability, adjust attitude, maneuver between trees and control landing. A morphology that produces high lift may be beneficial for extending glide distance or supporting body weight, but may also introduce higher drag or stronger pitching effects. Conversely, a morphology with lower peak lift but smoother post-stall behavior may provide more predictable aerodynamic response during maneuvering or landing. The present results do not directly test animal behavior, but they provide a mechanistic basis for linking membrane outline to aerodynamic function.

The tail and uropatagium simulations show a similar functional trade-off. The colugo-like T3 configuration generated stronger lift enhancement and pitching moment under symmetric deflection, suggesting that an integrated uropatagium can act as an effective posterior lifting and longitudinal-control surface. In contrast, the flat-tail T1 configuration generated stronger rolling and yawing moments under asymmetric deflection, suggesting a greater potential for lateral control. T2 showed intermediate behavior between these two cases. These results indicate that tail morphology may influence whether a gliding body is better suited for lift and pitch modulation or for lateral maneuvering.

The morphological data in Figure~\ref{fig5} support the idea that tail morphology varies substantially across gliding mammals. Although tail length generally increases with head--body length, the variation around the fitted trend indicates that body size alone does not explain tail form. Different groups of gliding mammals exhibit different combinations of tail length, membrane integration and body size. The CFD results suggest that such variation may have aerodynamic consequences, although ecological and evolutionary interpretations require further biological data.

From an engineering perspective, these findings provide design implications for bioinspired morphing aerial robots. In conventional aircraft, lifting surfaces and control surfaces are often treated as separate components. In contrast, gliding mammals use continuous membrane structures in which the limbs, patagium, tail and uropatagium interact aerodynamically. The present results suggest that morphology itself can be used as a design variable. A broader patagium may enhance lift, while a smoother pressure distribution may improve post-stall robustness. Similarly, an integrated uropatagium may improve lift and pitch control, while a distinct flat tail may enhance lateral maneuverability.

These principles are relevant for small aerial robots that operate at low Reynolds numbers and in cluttered environments. Such robots may benefit from lightweight membrane structures that combine load-bearing and control functions. Instead of relying only on conventional rigid control surfaces, a bioinspired robot could use patagium shape, tail integration or membrane deflection to tune its aerodynamic behavior. The selection of morphology would then depend on the desired function: glide efficiency, pitch control, roll/yaw maneuverability or robust post-stall behavior.

Several limitations should be noted. First, the simulations used simplified rigid geometries. Real gliding mammals have compliant membranes, flexible limbs, fur-covered surfaces and actively controlled body postures. Under aerodynamic loading, the patagium can deform and form camber, which may increase lift, delay stall and modify the pressure distribution. The present study intentionally neglected these aeroelastic effects to isolate the influence of planform geometry. Therefore, the results should be interpreted as a first-order comparison of morphological effects rather than a complete representation of animal gliding.

Second, the simulations were performed using a steady RANS approach with a realizable \(k\)-\(\varepsilon\) turbulence model. This approach is suitable for efficient comparative simulations across many geometries and angles of attack, but it has known limitations for separated and post-stall flows. In particular, unsteady vortex shedding, laminar separation bubbles and transient flow--membrane interactions may not be fully resolved. These limitations are most relevant at high angles of attack, where the flow is strongly separated. Future studies using higher-fidelity unsteady simulations, such as LES or hybrid RANS--LES methods, would be valuable for examining the detailed vortex dynamics and post-stall behavior.

Third, the pitching, rolling and yawing moment coefficients were evaluated about a prescribed reference point. This allows consistent comparison among the simulated morphologies, but it does not constitute a full static-stability analysis. In real animals and robots, the centre of gravity can vary with body posture, limb position and internal mass distribution, such as batteries, actuators and payloads. As a result, the absolute stability margin and control authority may change with CG location. Future robotic studies should therefore combine aerodynamic moment data with mass distribution and flight dynamics models.

Fourth, the simulations were conducted at a single free-stream velocity and Reynolds number. Although the selected condition is representative of the considered gliding-mammal-inspired model, smaller animals and micro aerial vehicles may operate at lower Reynolds numbers, where separation, transition and control-surface effectiveness can differ substantially. Future work should evaluate the robustness of the observed aerodynamic trends across a broader Reynolds-number range.

Finally, the present study does not include direct wind-tunnel validation for each isolated patagium and tail/uropatagium configuration. Experimental validation would be valuable, especially near stall and in the post-stall regime where CFD predictions can be sensitive to turbulence modeling and mesh resolution. Wind-tunnel tests using physical models with controlled membrane stiffness and tail deflection could further evaluate the effects of compliance and fabrication constraints.

In conclusion, this study shows that gliding-mammal-inspired patagium and tail/uropatagium morphologies produce distinct aerodynamic trade-offs. Patagium shape affects lift generation, drag, stall behavior and pitching moment, while tail and uropatagium morphology modulate longitudinal and lateral control authority. These results suggest that the diversity of gliding mammal morphology may reflect multiple aerodynamic functions rather than a single optimal design. They also provide design guidance for bioinspired morphing aerial robots that exploit membrane and tail morphology for aerodynamic performance and control.



\section*{Data availability statement}

The data that support the findings of this study are available from the corresponding author upon reasonable request.

\section*{Author contributions}

Liming Zheng: Conceptualization, Methodology, Investigation, Visualization, Writing -- original draft.
Baihui Chen: Methodology, Simulation, Data curation, Visualization, Writing -- review and editing.
Alexander van Zuijlen: Supervision, Methodology, Writing -- review.
Salua Hamaza: Supervision, Writing -- review.

\bibliographystyle{unsrt}
\bibliography{references}

@article{science_bat,
author = {F. T. Muijres  and L. C. Johansson  and R. Barfield  and M. Wolf  and G. R. Spedding  and A. Hedenström },
title = {Leading-Edge Vortex Improves Lift in Slow-Flying Bats},
journal = {Science},
volume = {319},
number = {5867},
pages = {1250-1253},
year = {2008},
doi = {10.1126/science.1153019},
URL = {https://www.science.org/doi/abs/10.1126/science.1153019},
eprint = {https://www.science.org/doi/pdf/10.1126/science.1153019}}

@book{book-CLCD,
  title={Mechanics of flight},
  author={Phillips, Warren F},
  year={2004},
  publisher={John Wiley \& Sons}
}

@article{morphing_wing_xr,
  title={Seamless active morphing wing simultaneous gust and maneuver load alleviation},
  author={Wang, Xuerui and Mkhoyan, Tigran and Mkhoyan, Iren and De Breuker, Roeland},
  journal={Journal of Guidance, Control, and Dynamics},
  volume={44},
  number={9},
  pages={1649--1662},
  year={2021},
  publisher={American Institute of Aeronautics and Astronautics}
}

@article{bookGlidingMammals,
  title={Gliding mammals of the world},
  author={Jackson, S},
  journal={CSIRP Pub},
  volume={156},
  year={2012}
}

@article{bio-fs-high-aoa,
  title={The relationship between 3-D kinematics and gliding performance in the southern flying squirrel, Glaucomys volans},
  author={Bishop, Kristin L},
  journal={Journal of Experimental Biology},
  volume={209},
  number={4},
  pages={689--701},
  year={2006},
  publisher={Company of Biologists}
}

@article{bio-fs-perching-trajectory,
  title={Glide performance and aerodynamics of non-equilibrium glides in northern flying squirrels (Glaucomys sabrinus)},
  author={Bahlman, Joseph W and Swartz, Sharon M and Riskin, Daniel K and Breuer, Kenneth S},
  journal={Journal of the Royal Society interface},
  volume={10},
  number={80},
  pages={20120794},
  year={2013},
  publisher={The Royal Society}
}

@article{bio-non-equilibrium-gliding,
  title={Global dynamics of non-equilibrium gliding in animals},
  author={Yeaton, Isaac J and Socha, John J and Ross, Shane D},
  journal={Bioinspiration \& Biomimetics},
  volume={12},
  number={2},
  pages={026013},
  year={2017},
  publisher={IOP Publishing}
}

@article{colugo1,
  title={Take-off and landing kinetics of a free-ranging gliding mammal, the Malayan colugo (Galeopterus variegatus)},
  author={Byrnes, Greg and Lim, Norman T-L and Spence, Andrew J},
  journal={Proceedings of the Royal Society B: Biological Sciences},
  volume={275},
  number={1638},
  pages={1007--1013},
  year={2008},
  publisher={The Royal Society}
}

@article{beihang-ieee-access,
  title={Aerodynamic characteristics and pitching adjusting mechanism of the flying squirrel with deployed patagium},
  author={Zhao, Fei and Wang, Wei and Zhang, Jingtao and Wyrwa, Justyna and Sun, Feng},
  journal={IEEE Access},
  volume={7},
  pages={185554--185564},
  year={2019},
  publisher={IEEE}
}

@article{colugo2,
  title={Sex differences in the locomotor ecology of a gliding mammal, the Malayan colugo (Galeopterus variegatus)},
  author={Byrnes, Greg and Lim, Norman T-L and Yeong, Charlene and Spence, Andrew J},
  journal={Journal of Mammalogy},
  volume={92},
  number={2},
  pages={444--451},
  year={2011},
  publisher={American Society of Mammalogists 810 East 10th Street, PO Box 1897, Lawrence~…}
}

@article{bio-red-giant-fs,
  title={Reproductive biology of the red-giant flying squirrel, Petaurista petaurista, in Taiwan},
  author={Lee, Pei-Fen and Lin, Yao-Sung and Progulske, Donald R},
  journal={Journal of mammalogy},
  volume={74},
  number={4},
  pages={982--989},
  year={1993},
  publisher={American Society of Mammalogists}
}

@article{robot-sr2025-squirrel,
  title={Monopedal robot branch-to-branch leaping and landing inspired by squirrel balance control},
  author={Yim, Justin K and Wang, Eric K and Lee, Sebastian D and Hunt, Nathaniel H and Full, Robert J and Fearing, Ronald S},
  journal={Science Robotics},
  volume={10},
  number={100},
  pages={eadq1949},
  year={2025},
  publisher={American Association for the Advancement of Science}
}

@article{robot-nc-Liming,
  title={A squirrel-inspired drone with enhanced stability, agility and maneuverability via whole-body morphing},
  author={Zheng, Liming and van Zuijlen, Alexander and Hamaza, Salua},
  journal={Nature Communications},
  year={2026},
  doi={10.1038/s41467-026-72822-w},
  publisher={Nature Portfolio}
}

@article{robot-Lentink-sr2,
  title={Bird-inspired reflexive morphing enables rudderless flight},
  author={Chang, Eric and Chin, Diana D and Lentink, David},
  journal={Science Robotics},
  volume={9},
  number={96},
  pages={eado4535},
  year={2024},
  publisher={American Association for the Advancement of Science}
}

@article{robot-Lentink-sr1,
  title={Soft biohybrid morphing wings with feathers underactuated by wrist and finger motion},
  author={Chang, Eric and Matloff, Laura Y and Stowers, Amanda K and Lentink, David},
  journal={Science Robotics},
  volume={5},
  number={38},
  pages={eaay1246},
  year={2020},
  publisher={American Association for the Advancement of Science}
}

@article{robot-Dario-sr1,
  title={Bioinspired wing and tail morphing extends drone flight capabilities},
  author={Ajanic, Enrico and Feroskhan, Mir and Mintchev, Stefano and Noca, Flavio and Floreano, Dario},
  journal={Science Robotics},
  volume={5},
  number={47},
  pages={eabc2897},
  year={2020},
  publisher={American Association for the Advancement of Science}
}

@article{robot-Dario-sr2,
  title={A twist of the tail in turning maneuvers of bird-inspired drones},
  author={Phan, Hoang-Vu and Floreano, Dario},
  journal={Science Robotics},
  volume={9},
  number={96},
  pages={eado3890},
  year={2024},
  publisher={American Association for the Advancement of Science}
}

@article{cfd-1,
  title={CFD Simulation of Flow Past MAV Wings.},
  author={Shetty, Pradeep and Subrahmanya, MB and Kulkarni, DS and Rajani, BN},
  journal={International Journal of Aerospace Innovations},
  volume={5},
  number={1},
  year={2013}
}

@article{cfd-3,
  title={XFOIL vs CFD performance predictions for high lift low Reynolds number airfoils},
  author={Morgado, JCPJ and Vizinho, R and Silvestre, MAR and P{\'a}scoa, JC},
  journal={Aerospace Science and Technology},
  volume={52},
  pages={207--214},
  year={2016},
  publisher={Elsevier}
}

@article{cfd-4,
  title={Numerical investigation of low-aspect-ratio wings at low Reynolds numbers},
  author={Cosyn, Peter and Vierendeels, Jan},
  journal={Journal of Aircraft},
  volume={43},
  number={3},
  pages={713--722},
  year={2006}
}

\end{document}